\documentclass{article}
\usepackage{amsfonts}

\usepackage{graphicx}
\usepackage{amsmath}

\newtheorem{theorem}{Theorem}
\newtheorem{acknowledgement}[theorem]{Acknowledgement}

\input{tcilatex}

\begin{document}

\title{Optimal ensemble length of mixed separable states}
\author{Robert B. Lockhart \\
Mathematics Department, United States Naval Academy, Annapolis MD 21402\\
\ rbl@nadn.navy.mil}
\maketitle

\begin{abstract}
The optimal (pure state) ensemble length of a separable state, A, is the
minimum number of (pure) product states needed in convex combination to
construct A. We study the set of all separable states with optimal (pure
state) ensemble length equal to $k$ or fewer. Lower bounds on $k$ are found
below which these sets have measure 0 in the set of separable states. In the
bipartite case and the multiparticle case where one of the particles has
significantly more quantum numbers than the rest, the lower bounds are
sharp. A consequence of our results is that for all two particle systems,
except possibly those with a qubit and those with a nine dimensional Hilbert
space, and for all systems with more than two particles the optimal pure
state ensemble length for a randomly picked separable state is with
probability 1 greater than the state's rank. In bipartite systems with each
particle having the same Hilbert space, with probablity 1 it is greater than
1/4 the rank raised to the 3/2 power and in a system with p qubits with
probability 1 it is greater than $2^{2p}/(1+2p)$, which is almost the
maximal rank squared.
\end{abstract}

\section{\protect\bigskip Introduction}

\bigskip

One of the important mathematical problems in quantum information theory is
the characterization of separable states. In the case of pure separable
states, much progress has been made. For instance, if one considers a
quantum system of $p$ particles with state space $H=\otimes _{j=1}^{p}%
\mathbb{C}^{n_{j}}$, then the pure states are rays in $H$. Mathematically,
this is the complex projective space$\mathbb{\ CP}\left( N-1\right) $, which
is a real manifold of dimension $2N-2$, where $N=n_{1}\cdots n_{p}$. The
separable pure states are product pure states and so correspond to a
submanifold isomorphic to the Cartesian product, $\mathbb{CP}\left(
n_{1}-1\right) \times \cdots \times \mathbb{CP}\left( n_{p}-1\right) $,
which has real dimension $\sum_{j=1}^{p}\left( 2n_{j}-2\right) $. Thus the
set of separable pure states is a measure 0, closed, non-dense subset of the
set of pure states. In particular if one randomly picks a pure state in $H$,
the probability it is entangled (i.e. not separable) is one. Moreover, every
entangled state has an open set of entangled states around it.

The situation for separable mixed states is quite different. To see why,
first recall that mixed states are described in terms of density matrices.
These are $N\times N$, complex, positive semi-definite, Hermitian matrices
with trace equal to $1$. If $N=n_{1}\cdots n_{p}$, then the separable
density matrices are those which are convex combinations of product
matrices, where by product matrix we mean one of the form $A=A_{1}\otimes
\cdots \otimes A_{p}$. Unlike the pure state case, the set of separable
density matrices, $\Sigma \left( n_{1},\ldots ,n_{p}\right) $, is not of
measure 0 in the set of all density matrices, $\mathcal{DM}\left( N\right) $
- it is not negligible. In fact the vector space of $N\times N$ Hermitian
matrices has bases which consist solely of product density matrices. This
means $\Sigma \left( n_{1},\ldots ,n_{p}\right) $ contains an open subset of 
$\mathcal{DM}\left( N\right) $, since the convex hull of a vector space
basis contains a set which is open in the hyperplane that contains the basis
elements. In the case of $\Sigma \left( n_{1},\ldots ,n_{p}\right) $, that
hyperplane is the set of matrices with trace equal to $1$. Thus $\Sigma
\left( n_{1},\ldots ,n_{p}\right) $ is a compact, convex subset of $\mathcal{%
DM}\left( N\right) ,$which is the closure of its non-empty interior. The
interior, moreover, contains an element which is in some sense the center of 
$\mathcal{DM}\left( N\right) $, the totally mixed state ( \cite{horodecki} 
\cite{braunstein} \cite{vidal} ).

One might think that $\Sigma \left( n_{1},\ldots ,n_{p}\right) $ would thus
be easy to characterize. After all, such is the case for common convex,
compact sets with non-empty interiors such as balls and polytopes. But $%
\Sigma \left( n_{1},\ldots ,n_{p}\right) $ is not simple at all. For
instance, unlike balls and polytopes, there is no easy way to determine the
minimum number of product states needed in convex combination to construct a
given separable mixed state.

If $A\in \Sigma \left( n_{1},\ldots ,n_{p}\right) $, we say its optimal
ensemble length is the minimum number of product states needed in convex
combination to construct $A$. When we require all the product states to be
pure, we call the minimum number needed the optimal pure state ensemble
length. This latter quantity was studied for two particle systems with $H=%
\mathbb{C}^{n}\otimes \mathbb{C}^{n}$ by Ulhmann \cite{uhlmann} and by
DiVincenzo, Terhal and Thapliyal \cite{divencenzo} among others. Uhlmann
showed the optimal pure state ensemble length is at least equal to the rank
of the density matrix and no greater than its square. DiVincenzo, Terhal and
Thapliyal took up the question of whether one actually needed more than the
rank. This is an important question, for the spectral theorem assures that
every density matrix can be expressed as the convex combination of pure
states, the number equalling the rank of the matrix. They found examples of
states with optimal pure state ensemble length greater than their rank. We
shall see for systems with three or more particles and for systems of two
particles other than \ possibly those modelled on $\mathbb{C}^{2}\otimes 
\mathbb{C}^{n}$ or $\mathbb{C}^{3}\otimes \mathbb{C}^{3}$ that almost every
separable state has an optimal pure state ensemble length greater than its
rank.

In this paper we examine the size of the set of all separable mixed states
which have optimal ensemble length of $k$ or fewer and the set of all which
have optimal pure state ensemble length of $k$ or fewer. The first set will
be denoted by $\Sigma ^{k}\left( n_{1},\ldots ,n_{p}\right) $ and the second
by $\Sigma _{pure}^{k}\left( n_{1},\ldots ,n_{p}\right) $. We completely
determine the $k$ for which $\Sigma ^{k}\left( n_{1},\ldots ,n_{p}\right) $
has measure 0 in $\Sigma \left( n_{1},\ldots ,n_{p}\right) $ in both the
bipartite case and the case in which one of the particles has substantially
more quantum numbers than all the rest - for instance a molecule and
photons. This result is the content of theorem 1. In theorem 2 (respectively
theorem 3) a lower bound on $k$ for which $\Sigma ^{k}\left( n_{1},\ldots
,n_{p}\right) $ (respectively $\Sigma _{pure}^{k}\left( n_{1},\ldots
,n_{p}\right) $) has measure 0 in $\Sigma \left( n_{1},\ldots ,n_{p}\right) $
is given. Moreover, in theorem 2 an upper bound on $k$ for which $\Sigma
\left( n_{1},\ldots ,n_{p}\right) $ has positive measure and contains an
open subset is also given. In order to put the main theorems in context, I
should mention that a classical theorem of Caratheodory assures one never
needs more than $N^{2}$ pure product states to construct a separable state.
Thus $\Sigma \left( n_{1},\ldots ,n_{p}\right) =\Sigma ^{k}\left(
n_{1},\ldots ,n_{p}\right) =\Sigma _{pure}^{k}\left( n_{1},\ldots
,n_{p}\right) $ for $k=N^{2}$. However, it is not the case one always needs
this many. For instance Sanpera, Tarrach and Vidral \cite{sanpera} have
shown in the 2-qubit case one needs no more than four pure product states.

Our main results are the following:

\begin{theorem}
\label{theorem 1} Let $N=n_{1}\cdots n_{p}$ with $n_{1}\leq n_{2}\leq \cdots
\leq n_{p}$ and $n_{1}\cdots n_{p-1}\leq n_{p.}$Then $\Sigma ^{k}\left(
n_{1},\ldots ,n_{p}\right) $ has the following properties: a) It is a
connected, compact subset of $\Sigma \left( n_{1},\ldots ,n_{p}\right) $. In
particular if it is not all of $\Sigma \left( n_{1},\ldots ,n_{p}\right) $,
then it is not dense and its complement in $\Sigma \left( n_{1},\ldots
,n_{p}\right) $ is an open subset . b) If $k<n_{1}^{2}\cdots n_{p-1}^{2}$,
then $\Sigma ^{k}\left( n_{1,}\ldots ,n_{p}\right) $ has measure 0 in $%
\Sigma \left( n_{1},\ldots ,n_{p}\right) $. c) If $k\geq n_{1}^{2}\cdots
n_{p-1}^{2}$, then $\Sigma ^{k}\left( n_{1},\ldots ,n_{p}\right) $ has
positive \ measure in $\Sigma \left( n_{1},\ldots ,n_{p}\right) $ and in
fact contains an open subset
\end{theorem}

\begin{theorem}
\label{theorem 2} The set $\Sigma ^{k}\left( n_{1},\ldots ,n_{p}\right) $
has the following properties: a) It is a connected, compact subset of $%
\Sigma \left( n_{1},\ldots ,n_{p}\right) $. In particular if it is not all
of $\Sigma \left( n_{1},\ldots ,n_{p}\right) $, then it is not dense and its
complement in $\Sigma \left( n_{1},\ldots ,n_{p}\right) $ is an open subset.
b) If $k<\left( n_{1}^{2}\cdots n_{p}^{2}\right)
/(1-p+\sum_{j=1}^{p}n_{i}^{2})$, then $\Sigma ^{k}\left( n_{1},\ldots
,n_{p}\right) $ has measure 0 in $\Sigma \left( n_{1},\ldots ,n_{p}\right) $%
. c) If $n_{1}\leq n_{2}\leq \cdots \leq n_{p}$ and $k\geq n_{1}^{2}\cdots
n_{p-1}^{2}$, then $\Sigma ^{k}\left( n_{1},\ldots ,n_{p}\right) $ has
positive measure in $\Sigma \left( n_{1},\ldots ,n_{p}\right) $ and in fact
contains an open subset.\ \ \ \ \ \ \ \ \ \ \ \ \ \ \ \ \ \ \ \ \ 
\end{theorem}

\begin{theorem}
\bigskip \label{theorem 3} The set $\Sigma _{pure}^{k}\left( n_{1},\ldots
,n_{p}\right) $ has the following properties: a) It is a connected, compact
subset of $\Sigma \left( n_{1},\ldots ,n_{p}\right) $. In particular if it
is not all of $\Sigma \left( n_{1},\ldots ,n_{p}\right) $, then it is not
dense and its complement in $\Sigma \left( n_{1},\ldots ,n_{p}\right) $ is
an open subset. b) If $k<\left( n_{1}^{2}\cdots n_{p}^{2}\right) /\left(
1-2p+\sum_{j=1}^{p}2n_{j}\right) $, then $\Sigma _{pure}^{k}\left(
n_{1},\ldots ,n_{p}\right) $ has measure $0$ in $\Sigma \left( n_{1},\ldots
,n_{p}\right) .$
\end{theorem}

The proofs of these theorems will be presented in the next section. First
though let us look at a few consequences.

In the bipartite case considered by Uhlmann and by DiVincenzo \textit{%
et.al., }$H=\mathbb{C}^{n}\otimes \mathbb{C}^{n}$. By theorem 1 there is an
open set of separable mixed states with optimal ensemble length of $n^{2}$
or fewer. Note $n^{2}=R_{\max }$, the maximal rank of density matrices in
this case. By theorem 3, however, the set of separable mixed states with
optimal pure state ensemble length equal to $n^{3}/\left( 4-3/n\right) $ or
fewer is of measure 0. Thus one must almost always use more than $\left(
R_{\max }\right) ^{3/2}/4$ pure product states to construct a mixed
separable state in the bipartite case. This is quite a disparity. However,
it is not indicative of all situations.

For instance, consider a system consisting of p qubits. Caratheodory's
theorem assures that every separable state can be decomposed into a convex
combination of $2^{2p}$ pure product states or fewer. From our theorems 2
and 3, it is seen that for large values of $\ p$ one must almost always use
close to that number, whether one uses pure product states or general ones.
In particular, $\Sigma ^{k}\left( 2,\ldots ,2\right) $ has measure 0 for $%
k<2^{2p}/(1+3p)$ and $\Sigma _{pure}^{k}\left( 2,\ldots ,2\right) $ has
measure 0 for $k<2^{2p}/(1+2p).$ In terms of the maximal rank, these
inequalities are $k<R_{\max }^{2}/(1+3\log (R_{\max }))$ and $k<R_{\max
}^{2}/(1+2\log (R_{\max }))$. On the other hand, theorem 2 implies $\Sigma
^{k}\left( 2,\ldots ,2\right) $ has positive measure and a non-empty
interior if $k\geq 2^{2p-2}$. This is not sharp. For instance, when $p=3$
one gets an open set with $k=13$.

Turning to the general multiparticle system, we note that the maximum rank
of a density matrix on $H=\mathbb{C}^{n_{1}}\otimes \cdots \otimes \mathbb{C}%
^{n_{p}}$ is $n_{1}\cdots n_{p}$. When this is less than $n_{1}^{2}\cdots
n_{p}^{2}/\left( 1-2p+\sum_{j=1}^{p}2n_{j}\right) $, we can conclude from
theorem 3 that the optimal pure state ensemble length of a separable state
is almost always greater than the rank of the state. In particular, one must
almost always use entangled pure states in the spectral (i.e. eigenvalue)
decomposition of separable states. This occurs for all systems with three or
more particles and for systems with two particles except possibly those with 
$H=\mathbb{C}^{2}\otimes \mathbb{C}^{n}$ or $H=\mathbb{C}^{3}\otimes \mathbb{%
C}^{3}$. That there are exceptions was shown by Sanpera \textit{et.at.} in 
\cite{sanpera}. As mentioned before, in that paper they showed every
separable state on $\mathbb{C}^{2}\otimes \mathbb{C}^{2}$ can be written as
the convex combination of four or fewer pure product states. It would be
interesting to see if the other $\mathbb{C}^{2}\otimes \mathbb{C}^{n}$ and $%
\mathbb{C}^{3}\otimes \mathbb{C}^{3}$ are also exceptions.

Before turning to the proofs, two things need to be mentioned about
measurability. First of all in this paper, ''almost always'' is used in the
strict mathematical sense of meaning ''except on a set of measure 0''.
Secondly, there is a controversy over the proper measure to use for the set
of density matrices. That does not apply to the results presented here since
any two measures which are absolutely continuous with respect to each other
have the same sets of measure 0. Since the hyperplane of Hermitian matrices
with trace equal to 1 is a real $N^{2}-1$ dimensional vector space and $%
\Sigma \left( n_{1},\ldots ,n_{p}\right) $ is a compact, convex subset of it
with non-empty interior, we shall use $N^{2}-1$ dimensional Lebesgue measure
for both.

\section{Proofs}

Suppose $M$ and $N$ are two finite dimensional $C^{\infty }$ manifolds and $%
f $ is a $C^{\infty }$ function from $M$ to $N$. A point $m\in M$ is a
critical point for $f$ if $df_{m}:TM_{m}\rightarrow TN_{f\left( m\right) }$
is not onto. In words: $m$ is a critical point for $f$ if the differential
of $f$ at $m$, which is a linear transformation from the tangent space of $M$
at $m$, $TM_{m}$, to the tangent space of $N$ at $f\left( m\right) $, $%
TN_{f\left( m\right) }$, is not onto. A point $n\in N$ is a critical value
for $f$ if it is the image of a critical point. A classical theorem in
differential topology due to Sard\cite{sard} states that the set of critical
values in $N$ is of measure $0$. This will be the key to our proofs. We
shall apply it, along with the rank theorem \cite{Narasimhan}, to the length 
$k$ mixing function which we shall define shortly.

For $w$ an integer, let $Herm\left( w\right) $ denote the set of $w\times w$
complex Hermitian matrices. $Herm\left( w\right) $ is a real vector space of
dimension $w^{2}$. The subset of positive semi-definite matrices in $%
Herm\left( w\right) $ form a closed, convex cone with non-empty interior.
For $r$ a real number take $\tau _{r}\left( w\right) $ to be the subset of $%
Herm\left( w\right) $ consisting of those matrices with trace equal to $r$.
Each $\tau _{r}\left( w\right) $ is a $w^{2}-1$ dimensional hyperplane in $%
Herm\left( w\right) $. They are all parallel to $\tau _{0}\left( w\right) $,
which is a vector space. The intersection of $\tau _{1}\left( w\right) $
with the cone of positive semi-definite matrices in $Herm\left( w\right) $
is the set of density matrices, $\mathcal{DM}\left( w\right) $. As mentioned
before, it is a compact, convex set with non-empty interior in $\tau
_{1}\left( w\right) $. Also, note that the tangent space of $\tau _{1}\left(
w\right) $ at $\mathbf{Q}$ is $\tau _{0}\left( w\right) $, since the
hyperplanes are parallel.

Let $N=n_{1}\cdots n_{p}$. The length $k$ mixing function 
\begin{equation*}
\mu _{k}:\mathbb{R}^{k-1}\times \left( \tau _{1}\left( n_{1}\right) \times
\cdots \times \tau _{1}\left( n_{p}\right) \right) ^{k}\rightarrow \tau
_{1}\left( N\right) 
\end{equation*}
is defined for $\mathbf{Q=}(\lambda _{1},\ldots ,\lambda
_{k-1},A_{11},\ldots ,A_{1p},\ldots ,A_{k1},\ldots ,A_{kp})$ by 
\begin{eqnarray*}
\mu _{k}\left( \mathbf{Q}\right)  &=& \\
&&\sum_{j=1}^{k-1}\lambda _{j}A_{j1}\otimes \cdots \otimes A_{jp}+\left(
1-\sum_{j=1}^{k-1}\lambda _{j}\right) A_{k1}\otimes \cdots \otimes A_{kp}
\end{eqnarray*}

When $\mu _{k}$ is restricted to $\Lambda _{k}\times \left( \mathcal{DM}%
\left( n_{1}\right) \times \cdots \times \mathcal{DM(}n_{p}\right) )^{k},$
where $\Lambda _{k}=\left\{ (\lambda _{1},\ldots ,\lambda _{k-1}):\lambda
_{j}\geq 0\text{ and }\sum_{j=1}^{k-1}\lambda _{j}\leq 1\right\} $, it
yields elements in $\mathcal{DM}(N)$. Moreover, it does so by forming convex
combinations of product states. Since $\mu _{k}$ is an algebraic function,
it is infinitely differentiable and so the criteria for Sard's theorem are
satisfied. The differential of $\mu _{k}$ at the point $\mathbf{Q}$ applied
to the tangent vector $\mathbf{V}=(r_{1},\ldots ,r_{k-1},H_{11},\ldots
,H_{kp})$ is given by:

\smallskip \smallskip 
\begin{eqnarray}
d\mu _{k}\left( \mathbf{Q}\right) \mathbf{V} &=&  \label{1} \\
&&\sum_{j=1}^{k-1}\lambda _{j}\left[ 
\begin{array}{c}
H_{j1}\otimes A_{j2}\otimes \cdots \otimes A_{jp}+A_{j1}\otimes
H_{j2}\otimes A_{j3}\otimes \cdots \otimes A_{jp} \\ 
\cdots +A_{j1}\otimes \cdots \otimes A_{jp-1}\otimes H_{jp}
\end{array}
\right]   \notag \\
&&+\left( 1-\sum_{j=1}^{k-1}\lambda _{j}\right) \left[ 
\begin{array}{c}
H_{k1}\otimes A_{k2}\otimes \cdots \otimes A_{kp}+\cdots  \\ 
+A_{k1}\otimes \cdots \otimes A_{kp-1}\otimes H_{kp}
\end{array}
\right]   \notag \\
&&+\sum_{j=1}^{k-1}r_{j}A_{j1}\otimes \cdots \otimes A_{jp}-\left(
\sum_{j=1}^{k-1}r_{j}\right) A_{k1}\otimes \cdots \otimes A_{kp}  \notag
\end{eqnarray}
\ \ \ \ \ \ \ 

We need to determine when $d\mu _{k}$ is never onto. To this end observe
that $\tau _{0}\left( N\right) $, the tangent space at each point of $\tau
_{1}\left( N\right) $, equals 
\begin{eqnarray}
&&\tau _{0}\left( n_{1}\right) \otimes Herm(n_{2})\otimes \cdots \otimes
Herm(n_{p})+  \label{2} \\
&&Herm\left( n_{1}\right) \otimes \tau _{0}\left( n_{2}\right) \otimes
\cdots \otimes Herm\left( n_{p}\right) +  \notag \\
&&\cdots +Herm\left( n_{1}\right) \otimes \cdots \otimes Herm\left(
n_{p-1}\right) \otimes \tau _{0}\left( n_{p}\right) .  \notag
\end{eqnarray}

(Note, this is sum, not direct sum. There is a great deal of overlap in the
terms. In particular, do not add dimensions.)

Let us first prove part b of theorem 1 for the bipartite case. Thus $%
N=n_{1}n_{2},$ $n_{1}\leq n_{2},$ and $k<n_{1}^{2}$. We need to show $d\mu
_{k}\left( \mathbf{Q}\right) $ is not onto for any $\mathbf{Q=}\left(
\lambda _{1},\ldots ,\lambda _{k-1},A_{11},A_{12},\ldots
,A_{k1},A_{k2}\right) $. To begin, notice that $k<n_{1}^{2}\leq n_{2}^{2}$
means that neither $\left\{ A_{j1}\right\} $ spans $Herm(n_{1})$ nor $%
\left\{ A_{j2}\right\} $ spans $Herm(n_{2}).$ Hence if either the
projections of the $A_{j1}$ onto $\tau _{0}\left( n_{1}\right) $ do not span 
$\tau _{0}\left( n_{1}\right) $ or the projections of the $A_{j2}$ onto $%
\tau _{0}\left( n_{2}\right) $ do not span $\tau _{0}\left( n_{2}\right) $,
then $d\mu _{k}\left( \mathbf{Q}\right) $ cannot be onto. Indeed, without
loss of generality suppose the projections of the $A_{j2}$ onto $\tau
_{0}\left( n_{2}\right) $ do not span. Then there is a $C\in \tau _{0}\left(
n_{2}\right) $ which is orthogonal to the span of those projections. Since $%
\left\{ A_{j1}\right\} $ does not span $Herm\left( n_{1}\right) $ there is a 
$B\in Herm\left( n_{1}\right) $ which is orthogonal to the span of $\left\{
A_{j1}\right\} $. The product $B\otimes C$ is then both in $\tau _{0}\left(
n_{1}n_{2}\right) $ and orthogonal to every term in equation (1) and so $%
d\mu _{k}\left( \mathbf{Q}\right) $ is not onto.

Since $\dim \tau _{0}\left( n_{2}\right) =n_{2}^{2}-1$, the situation just
considered occurs if any of the following hold $k<n_{1}^{2}-1,$ $n_{1}<n_{2}$%
, any of the $\lambda _{j}$ are 0, or the $\lambda _{j}$ add to 1.
Therefore, to finish this part of the proof let us assume $n_{1}=n_{2}=n$, $%
k=n^{2}-1,$ none of the $\lambda _{j}$ are 0 and the $\lambda _{j}$ do not
add to 1.

Suppose $A_{j1}=E_{j}+\frac{1}{n}I$ and $A_{j2}=F_{j}+\frac{1}{n}I$ where $%
\left\{ E_{j}\right\} $ and $\left\{ F_{j}\right\} $ are bases for $\tau
_{0}\left( n\right) $. In order to establish $d\mu _{k}\left( \mathbf{Q}%
\right) $ is not onto, we only need to show it does not send a basis of $%
\mathbb{R}^{k-1}\times \left( \tau _{0}\left( n\right) \times \tau
_{0}\left( n\right) \right) ^{k}$ onto a basis of $\tau _{0}\left(
n^{2}\right) .$ The elements of $\mathbb{R}^{k-1}\times \left( \tau
_{0}\left( n\right) \times \tau _{0}\left( n\right) \right) ^{k}$ are of the
form $\mathbf{V}=\left( r_{1},\ldots ,r_{k-1},H_{11},H_{12},\ldots
,H_{k1},H_{k2}\right) $. By successively picking one $r_{j}$ to be 1 and all
the other entries in $\mathbf{V}$ to be 0 and then picking all $r_{j}$ to be
0 and successively picking $H_{ji}$ to be one of the $E_{s}$ or $F_{t}$
depending upon whether $i=1$ or $2$, we obtain a basis for $\mathbb{R}%
^{k-1}\times \left( \tau _{0}\left( n\right) \times \tau _{0}\left( n\right)
\right) ^{k}$. Applying $d\mu _{k}\left( \mathbf{Q}\right) $ to this basis,
we obtain the set 
\begin{equation}
\left\{ 
\begin{array}{c}
E_{s}\otimes F_{t}+E_{s}\otimes \frac{1}{n}I,E_{s}\otimes F_{t}+\frac{1}{n}%
I\otimes F_{t}, \\ 
E_{s}\otimes F_{t}+\frac{1}{n}I\otimes F_{t}+E_{s}\otimes \frac{1}{n}%
I-E_{k}\otimes F_{k}-\frac{1}{n}I\otimes F_{k}-E_{k}\otimes \frac{1}{n}I
\end{array}
\right\}  \label{3}
\end{equation}
where $s$ and $t$ range independently from 1 to $n^{2}-1$. Subtracting the
first group of these elements from the second and third groups and adding
the first group with $s=t=k$ to the third, we get the set 
\begin{equation}
\left\{ E_{s}\otimes F_{t}+E_{s}\otimes \frac{1}{n}I,\frac{1}{n}I\otimes
F_{t}-E_{s}\otimes \frac{1}{n}I,\frac{1}{n}I\otimes F_{t}-\frac{1}{n}%
I\otimes F_{k}\right\}  \label{4}
\end{equation}
Subtracting the last group from the second and adding the result to the
first group, we obtain 
\begin{equation}
\left\{ E_{s}\otimes F_{t}+\frac{1}{n}I\otimes F_{k},\frac{1}{n}I\otimes
F_{k}-E_{s}\otimes \frac{1}{n}I,\frac{1}{n}I\otimes F_{t}-\frac{1}{n}%
I\otimes F_{k}\right\}  \label{5}
\end{equation}

Since $\dim \tau _{0}\left( n\right) =n^{2}-1$, there are $\left(
n^{2}-1\right) \left( n^{2}-1\right) $ elements in the first group of this
last set. There are $n^{2}-1$ elements in the second group and there are $%
n^{2}-2$ in the third group. Thus all told there are $n^{4}-2$ elements in
the set. But $\dim \tau _{0}\left( n^{2}\right) =n^{4}-1$ and so the set
cannot form a basis, which means $d\mu _{k}\left( \mathbf{Q}\right) $ is
never onto if $k<n_{1}^{2}$.

Hence if $k<n_{1}^{2}$, then every point in $\mathbb{R}^{k-1}\times \left(
\tau _{1}\left( n_{1}\right) \times \tau _{1}\left( n_{2}\right) \right)
^{k} $ is a critical point for $\mu _{k}$. It follows from Sard's theorem
that the image of $\mu _{k}$ is of measure 0 in $\tau _{1}\left( N\right) $.
The bipartite case of part b of theorem 1 is then a result of the facts that 
$\Sigma ^{k}\left( n_{1},n_{2}\right) $ is in the image of $\mu _{k}$ and
any measure $0$ subset of $\tau _{1}\left( N\right) $ has measure 0 in $%
\Sigma \left( n_{1},n_{2}\right) $ too.

To finish the proof of part b of theorem 1, we only need to note that $%
\Sigma ^{k}\left( n_{1},\ldots ,n_{p}\right) \subset \Sigma ^{k}\left(
\prod_{j=1}^{p-1}n_{j},n_{p}\right) $and use what we have just proved for
the bipartite case.

Let us now prove part c of theorems 1 and 2. We shall use the rank theorem 
\cite{Narasimhan} which states that if $d\mu _{k}$ is onto at a point $%
\mathbf{Q=}\left( \lambda _{1},\ldots ,\lambda _{k-1},A_{11},\ldots
,A_{kp}\right) $, then $\mu _{k}$ maps some open ball centered at $\mathbf{Q}
$ onto an open set containing $\mu _{k}\left( \mathbf{Q}\right) $. Thus we
need to find a  $\mathbf{Q}$ in the interior of $\Lambda _{k}\times \left( 
\mathcal{DM}\left( n_{1}\right) \times \cdots \times \mathcal{DM}\left(
n_{p}\right) \right) ^{k}$ at which $d\mu _{k}\left( \mathbf{Q}\right) $ is
onto.

We know there are bases of $Herm(n_{i})$ which consist of elements in the
interior of $\mathcal{DM}\left( n_{i}\right) $. We also know $\frac{1}{n_{p}}%
I$ is in the interior of $\mathcal{DM}\left( n_{i}\right) $. Therefore,
since $k\geq n_{1}^{2}\cdots n_{p-1}^{2}=\dim Herm\left( n_{1}\cdots
n_{p-1}\right) ,$ we can pick the $A_{ji}$ for $j=1,\ldots ,k,$ $i=1,\ldots
,p-1$ to be in the interior of $\mathcal{DM}\left( n_{i}\right) $ and such
that $\left\{ A_{j1}\otimes \cdots \otimes A_{jp-1}\right\} $ spans $%
Herm\left( n_{1}\cdots n_{p-1}\right) $. Choosing them so and also choosing
all $\lambda _{j}=\frac{1}{k}$ and all $A_{jp}=\frac{1}{n_{p}}I$, we obtain
a $\mathbf{Q}$ which satisfies our needs. To see this, note an element of $%
\mathbb{R}^{k-1}\times \left( \tau _{0}\left( n_{1}\right) \times \cdots
\times \tau _{0}(n_{p}\right) )^{k}$ is of the form $\mathbf{V}%
=(r_{1},\ldots ,r_{k-1},H_{11},\ldots ,H_{1p},\ldots ,H_{k1},\ldots ,H_{kp})$%
. Let $\Gamma _{i}$ be the set of all $\mathbf{V}$ for which the only
non-zero component is an $H_{ji}$. Since $\left\{ A_{j1}\otimes \cdots
\otimes A_{jp-1}\right\} $ spans $Herm\left( n_{1}\cdots n_{p-1}\right) $
and $A_{jp}=\frac{1}{n_{p}}I$, we have that $d\mu _{k}\left( \mathbf{Q}%
\right) $ maps $\Gamma _{i}$ onto $\tau _{0}\left( n_{1}\cdots n_{p}\right)
\otimes \frac{1}{n_{p}}I$ for $i<p$ and $\Gamma _{p}$ onto $Herm\left(
n_{1}\cdots n_{p-1}\right) \otimes \tau _{0}\left( n_{p}\right) $. These two
sets span $\tau _{0}\left( N\right) $ and so part c of theorems 1 and 2 is
proved.

To finish the proofs of these two theorems we first note that if $k$
satisfies the condition in part b of theorem 2, then the dimension of the
domain of $\mu _{k}$ is less than the dimension of its target. In such a
case it is impossible for $d\mu _{k}\left( \mathbf{Q}\right) $ to ever be
onto because the dimension of its domain is too small. And so the result
follows again from Sard's theorem. As for part a of theorems 1 and 2, it is
a consequence of the fact $\Sigma ^{k}\left( n_{1},\ldots ,n_{p}\right) $ is
the image of the connected, compact set $\Lambda _{k}\times \left( \mathcal{%
DM}\left( n_{1}\right) \times \cdots \times \mathcal{DM}\left( n_{p}\right)
\right) ^{k}$ under the continuous map $\mu _{k}$.

Finally, as for theorem 3, let us recall that the set of pure states in $%
\mathcal{DM}\left( q\right) $ is isomorphic to the complex projective space $%
\mathbb{CP}\left( q-1\right) $, which has real dimension $2q-2$. Hence we
need to consider the composition of the embedding

\begin{equation*}
\iota :\mathbb{R}^{k-1}\times \left( \mathbb{CP}\left( n_{1}-1\right) \times
\cdots \times \mathbb{CP}\left( n_{p}-1\right) \right) ^{k}\rightarrow 
\mathbb{R}^{k-1}\times \left( \tau _{1}\left( n_{1}\right) \times \cdots
\times \tau _{1}\left( n_{p}\right) \right) ^{k} 
\end{equation*}
with $\mu _{k}$. Part a of theorem 3 is a result of the fact $\Sigma
_{pure}^{k}\left( N\right) $ is the image of the connected, compact set $%
\Lambda _{k}\times \left( \mathbb{CP}\left( n_{1}-1\right) \times \cdots
\times \mathbb{CP}\left( n_{p}-1\right) \right) ^{k}$ under the continuous
map $\mu _{k}\circ \iota $. As for part b, it is a simple consequence of
Sard's theorem and the observation that $\mathbb{R}^{k-1}\times \left( 
\mathbb{CP}\left( n_{1}-1\right) \times \cdots \times \mathbb{CP}\left(
n_{p}-1\right) \right) ^{k}$ has dimension $k\left(
1-2p+\sum_{j=1}^{k}2n_{j}\right) -1$ while $\tau _{1}\left( N\right) $ has
dimension $n_{1}^{2}\cdots n_{p}^{2}-1$. Since $d\mu _{k}\circ \iota $ is a
linear transformation, it cannot be onto if the dimension of the domain is
strictly less than the dimension of its image, which is the case here if $%
k<\left( n_{1}^{2}\cdots n_{p}^{2}\right) /\left(
1-2p+\sum_{j=1}^{k}2n_{j}\right) $.

\begin{acknowledgement}
Research for this paper was partially funded by the Naval Research
Laboratory in Washington, D.C.
\end{acknowledgement}

\end{document}